# Deep Learning for GPS Spoofing Detection in Cellular-Enabled UAV Systems

Yongchao Dang *, Chafika Benzaïd *, Bin Yang *, and Tarik Taleb* †

* Aalto University, Espoo, Finland, †University of Oulu, Oulu, Finland

* Email: {yongchao.dang, chafika.benzaid, bin.1.yang, tarik.taleb}@aalto.fi

*Abstract*—Cellular-based Unmanned Aerial Vehicle (UAV) systems are a promising paradigm to provide reliable and fast Beyond Visual Line of Sight (BVLoS) communication services for UAV operations. However, such systems are facing a serious GPS spoofing threat for UAV's position. To enable safe and secure UAV navigation BVLoS, this paper proposes a cellular network assisted UAV position monitoring and anti-GPS spoofing system, where deep learning approach is used to live detect spoofed GPS positions. Specifically, the proposed system introduces a MultiLayer Perceptron (MLP) model which is trained on the statistical properties of path loss measurements collected from nearby base stations to decide the authenticity of the GPS position. Experiment results indicate the accuracy rate of detecting GPS spoofing under our proposed approach is more than 93% with three base stations and it can also reach 80% with only one base station.

*Index Terms*—UAV, GPS spoofing, Deep Learning, MLP, Path Loss.

## I. INTRODUCTION

Unmanned Aerial Vehicle (UAV) has received considerable scholarly attention in recent years, thanks to its high mobility, low cost, and flexible deployment in various civil and military fields, such as package delivery, precision agriculture, and aid communication [1]. Traditional UAVs have typically focused on the unlicensed spectrum (e.g., ISM 2.4 GHz) for communication and control in order to be free of charge and for ease of use. However, the unlicensed spectrum does not have a strong security mechanism and large-scale coverage ability, which restricts the UAV piloting Beyond Visual Line of Sight (BVLoS) [2].

More recently, cellular-enabled UAVs have been successfully armed and controlled remotely over 5G advanced facilities, which introduces a new formula to overcome the aforementioned shortcomings [3]. Nevertheless, the safe and secure navigation of UAVs is crucial for those remote operations. To that end, the Global Positioning System (GPS) is used by UAVs as a major navigation system to obtain their positions. However, the civil GPS uses unencrypted navigation signals and is vulnerable to spoofing attacks. Indeed, an attacker can use universal software radio peripheral (GPRS) to generate fake GPS satellite signals and fool the UAVs by obtaining a false position [4].

Several methods currently exist for the measurement of antiGPS spoofing that are mainly focusing on GPS navigation signals analysis [5]–[8], GPS navigation message authentication [9]–[12], Inertial Navigation System (INS) based spoofing detection [13]–[15], and Mobile Positioning System (MPS) based spoofing detection [16]–[18]. The GPS navigation signals analysis detects the spoofed signal by estimating and comparing the Direction of Arrival (DoA) of the GPS signal. The author in [5] introduced a multi-antennas to estimate the
DOA of GPS signals in order to verify the authenticity of the GPS. The work in [6] proposed a spatial signal processing approach for GPS spoofing detection and mitigation. The methods in [7] and [8] rely on the cross-correlation between encrypted/military GPS signals and civil GPS signals at the same position, where the encrypted GPS signals from the military are recognized as the trust temple for indicating the presence of a spoofing attack. Nevertheless, the adoption of those GPS signals analysis approaches either require multiantennas for estimating the DoA of GPS signal or needs a secure GPS receiver to perform the cross-correlation and incurs more computational load on the GPS receiver. Compared with GPS signal analysis, GPS navigation message authentication does not need more antennas or additional receivers. The GPS Navigation Message Authentication (NMA) approach protects the civil GPS signal from attacking by embedding the cryptographic signature into the navigation messages. Wu *et al.* presented a BeiDou-II NMA scheme based on digital signatures generated by Elliptic Curve Digital Signature Algorithm (ECDSA) in [9] and leveraged SM cryptographic algorithms to authenticate the BeiDou-II navigation messages in [10]. Wesson *et al.* [11] prevented counterfeit navigation messages by combining signature-based security methods with hypothesis tests. In [12], a trusted execution environment was used to generate cryptographically signed GPS messages in order to prevent their forgery. Nonetheless, even though NMA techniques are considered a practical and effective defense against GPS spoofing attacks, those techniques induce significant computational cost and latency due to signature verification. Inertial Navigation System (INS) techniques detect GPS spoofing by using the position information estimated from the Inertial Measurement Unit (IMU) that consisted of various onboard sensors including accelerometers, gyroscopes, magnetometers, and camera views, to cross-validate the veracity of the reported GPS position. Lee *et al.* [13] used accelerometer outputs and the acceleration computed from the GPS outputs to detect GPS spoofing. The authors in [14] used the UAV on-board gyroscopes' measurements to determine whether the GPS has been hijacked or not. In [15], the authors used the probability density function to analyze the accelerometer readings in order to identify spoofing GPS signals. Notwithstanding, the error accumulation of the IMU measurements is the main issue for INS, which can reduce the detection accuracy. Recently, Mobile Positioning System (MPS) based spoofing detection has emerged as a new class of anti-GPS spoofing approaches that leverages the localization ability of mobile cellular networks to relocate the UAV and discriminate the spoofed GPS positions in the base stations' coverage area. The work in [16] exploited the Receive Signals Strength (RSS) from 2G base stations to estimate the vehicle position and cross-check the vehicle's GPS position. In [17], the authors used the data relative to the neighboring cells to verify the GPS position. The authors in [18] considered the use of 5G network to infer the trust area where the GPS position should be located and recognize spoofed UAV's GPS positions. The MPS-based spoofing detection methods use the triangulation location technique, which requires at least three base stations at the same time for a desirable spoofing detection accuracy and is also sensitive to the environmental changes. Note that the GPS spoofing detection methods discussed above either depend on expensive hardware or can be negatively affected by environment changes. Therefore, these detection methods are difficult to be used in UAV systems due to the inherent characteristics of UAV, such as fast movement, limited storing and computing capacity.

To date, there have been very few empirically published accounts of an effective GPS spoofing detection approach that accommodates resource, cost, and environmental constraints. For this purpose, we propose a new solution to devise an effective cellular-enabled UAV GPS spoofing detection system, where a deep learning algorithm, specifically MLP, is used to analyze the statistical features of path losses between UAV and base stations (BSs). The proposed approach can conduct on the edge server without any additional hardware or computation load at the UAV. In addition, its effectiveness is less prone to changes in environmental conditions, thanks to the stability introduced by the statistical features. By using the path losses that can be obtained from the BSs broadly and speedily [19], and by taking advantage of the capability of ML to deliver faster decisions, the proposed approach will empower live detection of spoofed GPS positions. The main contributions of this paper are summarized as follows: . Firstly, we propose an UAV position monitoring and anti-GPS spoofing system, wherein the hypothesis testing compares the path loss measurements collected from the BSs and the associated theoretical path losses corresponding to the reported position to empower live detection of spoofed GPS positions.

- Secondly, in order to make the detection approach insensitive to changes in environmental conditions, the proposed MLP models use three statistical properties of path losses as inputs, including moments (e.g., Mean Variance Skewness Kurtosis (MVSK)), quartile (e.g., BOX), and probability distributions difference (e.g., Wasserstein Distance (WD)).

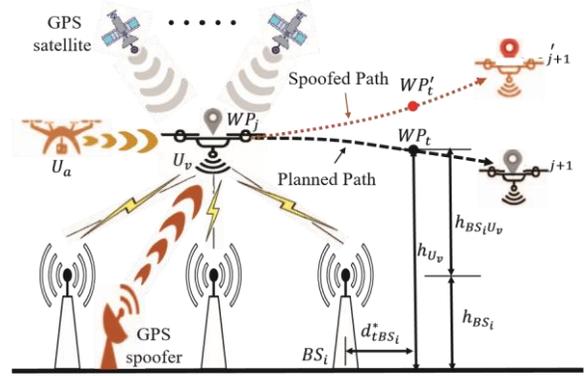

Fig. 1. Network model

- Thirdly, We then develop three MLP models, namely MVSK-MLP, BOX-MLP and WD-MLP, trained on the statistical properties to decide whether the reported GPS positions are fake or legitimate.
- Finally, the results of the simulation study demonstrates the effectiveness of the proposed MLP models in detecting the GPS spoofing attacks under different base stations scenarios.

The rest of this paper is organized as follows. The system model is described in Section II. Section III introduces three MLP models to detect GPS spoofing based on statistical properties of path losses. The performance of the MLP models are evaluated in Section IV. Section V concludes this paper.

## II. SYSTEM MODEL

This section describes the network and communication models considered in this study and defines the hypothesis testing method used to formulate the GPS spoofing detection problem.

*1) Network Model:* As illustrated in Fig. 1, we consider a network scenario consisting of a victim UAV $U_v$, an active malicious UAV $U_a$, GPS satellites, and $N$ BSs. The active malicious UAV can send fake GPS signals to the victim UAV. Let $(x_{BSi}, y_{BSi}, h_{BSi})$ denote the location of the $i$th BS. In absence of spoofing, $U_v$ should be located at time $t$ at waypoint $WP_t = (x_t, y_t, h_t)$ of the planned trajectory consisting of waypoints $WP_i$ and $WP_{i+1}$. Meanwhile, if we assume that the GPS spoofing starts when $U_v$ is at $WP_j = (x_j, y_j, h_j)$, its position at time $t$ will be at $WP_t$ of the spoofed path.

*2) Communication Model:* According to the 3GPP definition in [20], we use both Line-of-Sight (LoS) links and NonLine-of-Sight (NLoS) links to model the channel from UAV to BS. The theoretical path loss $\overline{PL_{BSiUv}}$ between $BS_i$ and $U_v$ is defined in 3GPP document in [20].

*3) Hypothesis Testing:* Based on the wireless signal attenuation theory, the path losses values between the base station and the UAV become bigger as the distance increases.

For different positions, we can observe the difference path loss values generally. In this paper, the actual path loss, $PL_{BS_iU_v}$, is provided by BSs, and the theoretical path loss, $\overline{PL_{BS_iU_v}}$, is determined by the UAV reported position according to the communication channel model (See Subsection II-2). Thus, we have

$$\Delta L_{BS_iU_v} = \left| PL_{BS_iU_v} - \overline{PL_{BS_iU_v}} \right| \quad (1)$$

where $\Delta L_{BS_iU_v}$ is the absolute difference between the actual path loss and the theoretical one. It is a widely held view that an actual GPS position of UAV corresponds to a theoretical path loss that is nearly the same as the actual one. On the contrary, a spoofed GPS position refers to a theoretical path loss deviating from the actual one. In other words, a bigger $\Delta L_{BS_iU_v}$ indicates that the GPS position of the UAV is spoofed with a higher probability. Hence, the GPS spoofing detection problem is formulated as a hypothesis testing given by

$$\begin{cases} H_0: & \Delta L_{BS_iU_v} > T, \\ H_1: & \Delta L_{BS_iU_v} \leq T, \end{cases} \quad (2)$$

where $T$ is a threshold of the hypothesis testing. The null hypothesis $H_0$ represents that the GPS position is spoofed. $H_0$ is accepted if $\Delta L_{BS_iU_v}$ is above the threshold $T$. On the other hand, a true alternative hypothesis $H_1$ is proposed for a higher probability of no GPS spoofing.

It is possible that the threshold of path losses difference in hypothesis testing does not represent the real distance deviation between UAV and BS, because the path loss value is not only decided by the distance between UAV and BS but also impacted by other environmental factors (e.g., cloud, temperate and vapor). Therefore, the threshold based hypothesis testing for GPS spoofing detection faces the following significant challenges. Firstly, the environment impact on the path loss value is more likely to result in increasing the error of spoofing detection. Secondly, an inappropriate threshold value can yield false alarms or may result in missed detection. In fact, a smaller threshold value increases the probability of false alarms, while a bigger threshold value leads to a higher probability of missed detection. Thus, deciding the apposite threshold value is a vital yet difficult task. In addition, the hypothesis testing results issued by different BSs should be assigned different weights to make the final decision. The motivation behind using different weights is that a larger distance between a BS and a UAV could lead to a higher error on the hypothesis testing result.

To overcome those challenges, we leverage the potential of both statistical methods and deep learning algorithms to devise an effective GPS spoofing detection approach for cellular-enabled UAVs. To remove the effects of the changing environmental conditions, three statistical methods are introduced to extract the statistical properties of path losses data in a given time slot. Furthermore, the Multi-layer Perceptron (MLP) algorithm is used to deal with the threshold and weight setting issues. In fact, the MLP algorithm brings the advantage of intelligently finding the best threshold and assigning weights for discriminating fraudulent GPS locations.

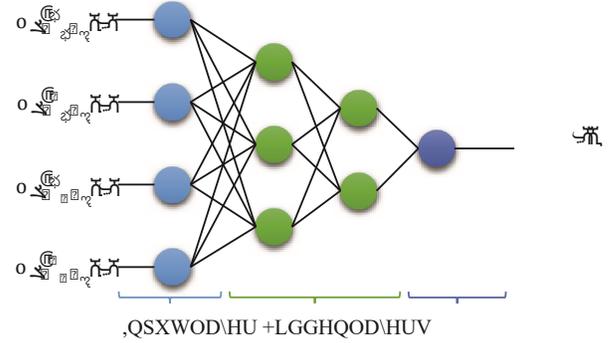

Fig. 2. The structure of the MLP.

### III. MLP-BASED GPS SPOOFING DETECTION MODEL

In this section, we first introduce the MLP model for GPS spoofing detection. Then, we build three models based on MLP algorithm using different statistical information extracted from the path losses data. As illustrated in Fig. 2, MLP is a deep learning neural network including an input layer, several hidden layers and an output layer, and each layer has an arbitrary number of neurons that propagate an output to the next layer through a nonlinear activation function. Mathematically, this can be formulated as

$$y = f\left(\sum_{j=1}^{n} \omega_j x_j + \theta\right) \quad (3)$$

where $x$ and $y$ denote the input vector and the output vector, respectively. $\omega$ is the weight vector and $\theta$ is the bias. $f(\cdot)$ is the nonlinear activation function.

The statistical properties are used as inputs to the devised MLP models. The input $\Delta L_{BS_nU_v}^{x_m}(t)$ denotes the $m$th difference of path losses reported by $BS_n$ under each of the statistical methods (i.e., MVSK [21], BOX [22], WD [23]). $x_m \in \{M,V,S,K\}$ for MVSK-based MLP (MVSK-MLP) model, $x_m \in \{Q_0,Q_1,Q_2,Q_3,Q_4\}$ for BOX-based MLP (BOX-MLP) model and $x_m = W$ for WD-based MLP (WD-MLP) model. The models' output, Prediction($t$), is the final prediction decision, i.e., the reported GPS position is spoofed or not. The three MLP models are then trained using a training Mean Squared Error (MSE) as performance indicator. The dataset **D** and evaluated with an unseen test dataset **D** using MSE is expressed as

$$MSE = \frac{1}{M}\sum_{i=1}^{M}(Y_i - \widehat{Y}_i)^2 \quad (4)$$

where $M$ represents the number of observations, $Y_i$ is the $i$th value.

predicted observation value, and $Y_i$ is the $i$th real observation

By leveraging the stability introduced by the statistical methods and the proven ability of MLP to learn any nonlinear and complex relationships between inputs and outputs, the proposed models are able to deliver accurate decisions. However, the models' efficiency in detecting GPS spoofing attack will depend greatly on the statistical metrics captured by the statistical methods. In fact, MVSK-MLP model requires a large amount of path loss data to ensure the accuracy of prediction. By removing outliers, the BOX-MLP model could mitigate the environment impacts on the path losses. Meanwhile, it will also lead to increased error in GPS spoofing detection since the outliers caused by attackers are also removed. The WD-MLP is used to describe the difference between actual and theoretical path losses, and thus generates only one feature value on the difference for each base station, which could result in unfitting problem.

## IV. PERFORMANCE EVALUATION

In this section, we assess the effectiveness of the proposed MLP-based GPS spoofing detection approach by comparing the performance of the three MLP models considering different number of BSs.

### A. Simulation Settings

To assess the performances of the proposed MLP models, we develop a simulator using Python 3.6 and Tensorflow 2.1. Python is used to set up the simulation platform and Tensorflow is applied to build the ML models. We consider three BSs, $BS_1, BS_2$ and $BS_3$, distributed at the fixed locations (0,0,35),(150,150,35) and (300,150,35) in a 3D space, respectively. Here, the BSs are 35 meters high. The evaluation is conducted under three network scenarios, namely the UAV is under the coverage of three, two or one BSs, respectively. The first scenario includes the three BSs and the second scenario includes $BS_1$ and $BS_3$. The second scenario can show the worst detection situation in our tests, as the distances between these BSs and the UAV start points/end points are relatively far. In addition, the third scenario including one BS $BS_1$ focuses on the average behavior of MLP-based GPS spoofing detector.

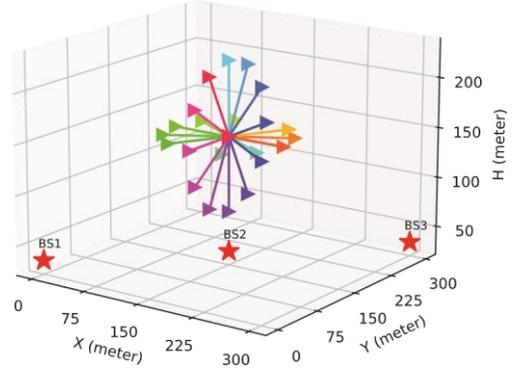

Fig. 3. Base stations and trajectory.

The UAV starts its mission from position (150,150,150) and flies to a destination 100 meters away from the starting location. To simulate the GPS spoofing attack, we consider 16 potential destinations evenly distributed over a 3D space as shown in Fig.3. One destination represents the real one while the others are spoofed. The communication links between the BSs and UAV obey the 3GPP definition in [20] and the channel frequency is set to 2.0 GHz. The value of $N$ used to extract the statistical metrics of path losses is set to 100. This allows to improve the detection accuracy while reducing the spoofing detection latency.

It is well known that the performance of MLP is sensitive to hyperparameter settings [24]. Thus, to find the best configuration of the different MLP models of our study, we carry out hyperparameter tuning by varying the learning rate (LR), the number of hidden layers, and the number of neurons per hidden layer. LR is drawn from {0.05,0.01,0.005,0.001,0.0005,0.0001}, the number of hidden layers is varied from 1 to 6 layers, and the number of neurons of each hidden layer is taken from {8,16,32,64,96,128}. The hidden layers have the same number of neurons and use Rectified Linear Unit (ReLU) as activation function. The MLP models are built using backpropagation method on a training dataset **D** containing 2259 samples and their prediction accuracy is evaluated on the unseen test dataset **D** comprising 969 samples. The training is performed for at most 500 epochs and an early stopping patience of 15 on MSE is applied to prevent overfitting. The model's MSE (see Eq.4) is calculated for every epoch on a held out validation set. Table.I summarizes the hyperparameters values of the best models produced for each MLP algorithm under each network scenario.

TABLE I
HYPERPARAMETERS SETTINGS OF THE BEST MLP MODELS.

| Scenario | Setting | MLP Algorithm | | |
|---|---|---|---|---|
| | | MVSK | BOX | WD |
| Three BSs | LR | 0.005 | 0.001 | 0.0005 |
| | Inputs | (12,0) | (15,0) | (3,0) |

504

|  | Hidden layers | 4 | 5 | 3 |
|---|---|---|---|---|
|  | Neurons | 96 | 96 | 16 |
| Two BSs | LR | 0.001 | 0.001 | 0.001 |
|  | Inputs | (8,0) | (10,0) | (2,0) |
|  | Hidden layers | 2 | 5 | 2 |
|  | Neurons | 32 | 96 | 64 |
| One BSs | LR | 0.0001 | 0.001 | 0.0001 |
|  | Inputs | (4,0) | (5,0) | (1,0) |
|  | Hidden layers | 2 | 5 | 2 |
|  | Neurons | 96 | 96 | 64 |

From Table I, it is noticed that different configurations are required for MLP models to reach their best performance under each of the three considered scenarios. The following key observations can be made: (i) the number of inputs to each MLP algorithm depends on the number of BSs involved in each scenario and the statistical method considered; (ii) the BOX-MLP algorithm needs more hidden layers and neurons compared to MVSK-MLP and WD-MLP algorithms. This can be explained by the fact that the number of hidden layers and neurons usually depends on the size of the input vector. In our case, BOX-MLP has the largest number of inputs; (iii) The WD-MLP algorithm uses at most 3 hidden layers and 64 neurons per hidden layer to achieve the best performance; (iv) Unlike MVSK-MLP and WD-MLP algorithms, WD-MLP algorithm requires the same MLP structure and LR to get the best performance for the three scenarios.

### B. Performance Results

*1) Performance metrics:* To assess the performance of the proposed MLP models for GPS spoofing detection, we use two performance metrics, namely MSE defined in Eq. (4), and *Accuracy* defined as

$$Accuracy = \frac{TP + TN}{TP + FP + FN + TN}, \quad (5)$$

where the *Accuracy* refers to the percentage of the reported GPS positions that are correctly classified. *TP* (True Positive) is the correctly detected spoofed positions, *FN* (False Negative) is the spoofed positions considered as normal positions, *FP* (False Positive) is the normal positions identified as spoofed, and *TN* (True Negative) is the normal positions that are correctly classified as normal.

*2) MLP Models Comparison Under Different Scenarios:* To investigate the performance of MVSK-MLP, BOX-MLP and WD-MLP models during their training, we record the accuracy and MSE after each training epoch. Fig. 4 and Fig. 5 show the results of the best models produced for each MLP algorithm under each network scenario.

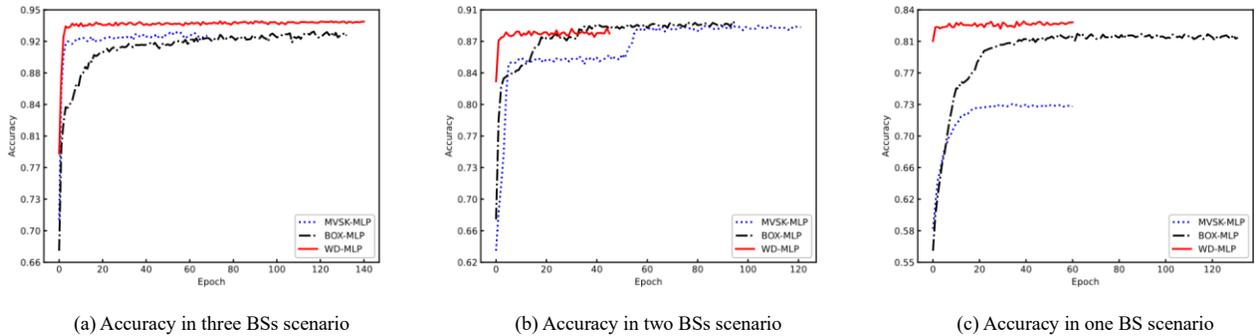

(a) Accuracy in three BSs scenario  (b) Accuracy in two BSs scenario  (c) Accuracy in one BS scenario

Fig. 4. Accuracy in different BSs scenarios

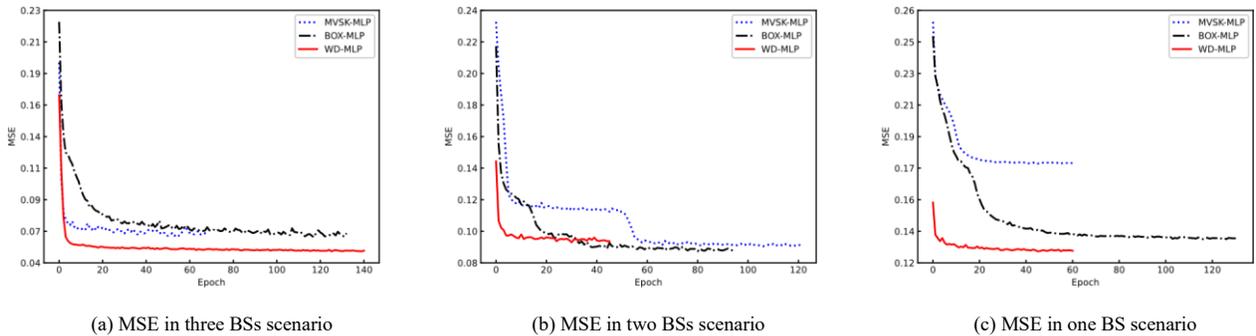

(a) MSE in three BSs scenario  (b) MSE in two BSs scenario  (c) MSE in one BS scenario

Fig. 5. MSE in different BSs scenarios

As shown in Fig. 4, the accuracy quickly increases after the first few epochs, after which the gain in accuracy is minimal before reaching a stable state. In Fig. 4(b), the accuracy trends of MVSK-MLP and BOX-MLP have a second steep around 55th epoch and 25th epoch, respectively, which indicates that the used stochastic gradient-based optimization method (i.e., Adam) is trapped in a local optimum between the two steeps. In Fig.4(c), the WD-MLP graph shows very minimal variations in accuracy over the epochs. This trend stems from the fact that only one WD value is used which allows the stochastic optimization method to easily find the best weights for WDMLP to reach the highest accuracy. From Fig. 4, we also observe that the number of BSs has a direct impact on the accuracy. The results show that the model accuracy is above 93% using three base stations and can reach 80% with only one base station. In fact, increasing the number of BSs will increase the amount of path losses data collected, which leads to improved detection accuracy. Furthermore, Fig. 4 shows that WD-MLP achieves the best detection accuracy for the three scenarios, which demonstrates that the statistical properties extracted by WD method adequately capture the changes in path losses. The results in Fig. 5 show that MSE has an opposite trend to accuracy. In fact, the MSE exhibits a downward trend over the training epochs for all scenarios and MLP models. It is clear that the improve in accuracy over the training epochs results in decrease in the prediction error.

## V. Conclusion

This paper proposed a cellular-enabled UAV position monitoring and anti-GPS spoofing system, wherein the path loss measurements collected from the base stations are used to cross-validate the UAV's GPS position. In the proposed system, we leveraged the potential of both statistical methods and deep learning to develop a novel MLP based GPS spoofing approach. The performance results demonstrated the effectiveness of the proposed approach in delivering accurate decisions, thanks to the stability introduced by the statistical metrics in enhancing the prediction accuracy. Indeed, the developed MLP approach could achieve an Accuracy rate that is above 93% with three base stations and can reach 80% with only one base station.

In the future, we will explore the potential of ensemble learning methods to further improve the spoofing detection accuracy by combining different MLP predictions. Furthermore, real data collected from UAV flying in real environment will be used to validate the envisioned MLP ensemble models.

## VI. Acknowledgement

This work was partially supported by the European Union's Horizon 2020 Research and Innovation Program through the 5G!Drones Project under Grant No. 857031, the Academy of Finland 6Genesis project under Grant No. 318927 and the INSPIRE-5Gplus project under Grant No. 871808. It was also supported in the National Outstanding Youth Science Fund Project of China with grant No. 61825104 and the National Natural Science Foundation of China under grant agreement No. 61941105.